\documentclass{he_symp}
\usepackage{psfig,graphicx,epsfig}
\usepackage{color}
\usepackage{amsmath,amssymb,epic,eepic,array}
\begin{document}
\renewcommand{\FirstPageOfPaper }{ 193}\renewcommand{\LastPageOfPaper }{ 199}
%
\providecommand{\VV}[1]{\mbox{\boldmath{$#1$}}}
\providecommand{\dif}[2]{\frac{\partial\protect#1}{\partial\protect#2}}%
\providecommand{\Dif}[2]{\frac{\mathrm{d}\/\protect#1}{\mathrm{d}\/\protect#2}}%
\providecommand{\nab}[0]{\VV{\nabla}}																		
\providecommand{\ro}[1]{{\rm{}#1}}																		
\providecommand{\Achtung}[1]{\emph{#1}\marginpar{\Huge\textbf{!}}}
\providecommand{\acro}[1]{\protect{}\textsc{#1}}
\providecommand{\Unit}[1]{\,\mbox{$\rm #1$}}
\providecommand{\sm}{{\mbox{\small{}-}}}
\providecommand{\wpe}{\ensuremath{\omega_\ro{pe}}}
\providecommand{\wpi}{\ensuremath{\omega_\ro{pi}}}
\providecommand{\te}{\ensuremath{T_\ro{e}}}
\providecommand{\tb}{\ensuremath{T_\ro{b}}}
\providecommand{\ti}{\ensuremath{T_\ro{i}}}
\providecommand{\wce}{\ensuremath{\omega_\ro{ce}}}
\providecommand{\wci}{\ensuremath{\omega_\ro{ci}}}
\providecommand{\vte}{\ensuremath{v_\ro{te}}}
\providecommand{\vti}{\ensuremath{v_\ro{ti}}}
\providecommand{\ld}{\ensuremath{\lambda_\ro{D}}}
\providecommand{\me}{\ensuremath{m_\ro{e}}}
\providecommand{\mi}{\ensuremath{m_\ro{i}}}
\providecommand{\kb}{\ensuremath{k_\ro{B}}}
\providecommand{\eps}{\ensuremath{\varepsilon_0}}
%
\title{The Free Electron Maser in Pulsar Magnetospheres}
\author{R.~Schopper\inst{1,4}, H.~Ruhl\inst{2}, T.A.~Kunzl\inst{3}
        and H.~Lesch\inst{4}} 
\institute{%
  Max--Planck--Institut f\"ur extraterrestrische Physik,
  Giessenbachstra{\ss}e, 85740 Garching, Germany 
\and  
  General Atomics, San Diego, CA, USA
\and
  Max Planck Institut f\"ur Quantenoptik, Garching, Germany
\and
  Universit\"ats--Sternwarte~M\"unchen, Scheinerstra\ss{}e~1,
  81679~M\"unchen, Germany, Centre for Interdisciplinary Plasma Science
}
\maketitle

\begin{abstract}
  We present the numerical simulations of coherent inverse Compton
  scattering (\acro{CICS}) in a highly magnetized plasma process by means
  of a full three dimensional particle in cell code~(\acro{PIC}), which is
  mass and energy conservative.  We used the parameters of a pulsar
  magnetosphere where \acro{CICS} is one of the most promising models for
  the generation of the observed highly coherent radio emission.  First we
  show details of the onset and time evolution of strong Langmuir
  turbulence driven by a relativistic electron beam penetrating a strongly
  magnetized background plasma.  The Langmuir turbulence acts as
  self-generated wiggler fields which bunch the beam electrons thereby
  inducing strong coherent emission of the bunches at frequency $\gamma^2$
  times the plasma frequency.  The emitted power is about $10 \Unit{GW}$ in
  a few nanoseconds.  This radiation is interpreted in terms of inverse
  Compton scattering on nonlinear density fluctuations. \acro{CICS} is the
  longitudinal version of a laboratory free electron laser and is
  applicable in strongly magnetized plasmas like pulsars.
\end{abstract}

\section{Introduction}
  A strong source for coherent radiation in laboratory is the free electron
  laser (\acro{FEL}) \cite{OShea}.  Its principle is the interaction of a
  relativistic electron beam (\acro{REB}) and a magnetic wiggler system.
  The beam of high energy electrons with energy $\gamma m_{\rm e} c^2$
  passes between permanent magnets of alternating polarity and periodicity
  $\lambda_{\rm W}$.  The energy of the \acro{REB} is efficiently converted
  by electron bunching.  In the observer's frame the bunches emit the
  relativistically Doppler-shifted wavelength $\lambda =\lambda_{\rm
  W}/\gamma^2$.  A comparable effect can be driven by a \acro{REB}
  penetrating a strongly magnetized plasma.  In phase space such a
  configuration corresponds to a two-stream-instability.  In the plasma
  version the free energy of the emitted radiation is supplied by the
  \acro{REB}, as it is in the case of the \acro{FEL}, but the positive
  gradient of the momentum distribution function $df/dp > 0$ is the source
  of the instability which excites strong density fluctuations.  In a
  strongly magnetized plasma such density oscillations are predominately
  longitudinal Langmuir waves which grow to nonlinear amplitudes
  \cite{Benford,Kato,Weatherall}.  The interaction of the \acro{REB} with
  the electrostatic fluctuations of the Langmuir waves results in very
  strong coherent electromagnetic waves \cite{Kato,Levron}.  In the rest
  frame of the beam electrons, they experience the Doppler shifted
  electrostatic wave moving towards them.  The nonlinear waves force the
  \acro{REB} to wiggle and to emit dipole radiation of the same frequency
  as the waves.  In the laboratory frame this Hertz' dipole radiation is
  again Doppler boosted and strongly beamed in forward direction due to the
  relativistic lighthouse effect.  This interaction can be viewed as
  coherent inverse Compton scattering (\acro{CICS}) off Langmuir waves
  \cite{Benford}.  Coherence occurs due to the bunch structure of the beam
  which means a phase coupled electrostatic field and density modulation.

\section{Computational Details}
  Since \acro{CICS} has been discussed to be promising candidate for the
  origin of the extremely coherent radio emission from pulsars
  \cite{Kunzl,Melrose1999,Melrose2000}, we use the physical parameters of a
  pulsar magnetosphere as an instructive application.  Radio observations
  indicate that at about 500 km from the strongly magnetized neutron star
  the coherent pulsed radio emission is produced.  The coherence is proved
  by pulse substructures on time scales down to a few tens of nanoseconds
  with increasing flux densities\cite{Rickett}.  In these regions of a
  pulsar magnetosphere typical values are: background electron
  density~$n_\ro{e} = 10^{12} \Unit{m}^{\sm 3}$, beam electron density
  equal to the plasma density~$n_\ro{b} = 10^{12} \Unit{m}^{\sm 3}$, beam
  electron Lorenz factor~$\gamma = \sqrt{5}$~($\gamma\beta = 2$) and an
  extremely strong background magnetic field~$B$ of about~$1000 \Unit{T}$.
  Especially the strong guiding field is important for the efficiency of
  the \acro{CICS}-process.  It reduces the electron dynamics to the spatial
  dimension along the magnetic field lines which greatly improves the
  stability of the excited Langmuir waves and thus leads to strong and
  stable electrostatic wiggler fields \cite{Pelletier}.  For the
  temperature of both the beam-- and background electrons values we take
  the value of~$\te = \tb = 100 \Unit{eV}$, typical polar cap temperature
  of neutron stars derived from their thermal X-ray-emission \cite{Pavlov}.
  The \acro{PIC}--code used for this numerical simulation is fully three
  dimensional and it conserves mass and energy.  In the simulation the
  direction of electron beam propagation~$\gamma\VV{\beta}$ and of the
  magnetic field~$\VV{B}$ is the positive $z$--direction, which is also
  called the longitudinal direction.  The simulated box has a numerical
  extension of $80 \times 80 \times 200$~grid points and a physical
  extension of ~$100 \Unit{m} \times 100 \Unit{m} \times 250 \Unit{m}$,
  which corresponds to the size of the expected features.  The resolution
  of $\delta x = \delta y = \delta z = 1.25 \Unit{m}$ is more than
  sufficient to resolve the expected maximum wavelengths of
  approximately~$\lambda_\ro{Rad} = 2\pi/k_\ro{Rad} = 2\pi c/\gamma^2 \wpe
  \approx 6.7 \Unit{m}$.  A particle density of $10^{12} \Unit{m}^{\sm 3}$
  is approximated by 10 quasi particles per cell, leading to 16 million
  background quasi electrons that are accompanied by a varying number of
  beam quasi electrons of not more than 2.5 million, which eventually enter
  the computational box.  Those numbers are doubled, since for quasi
  neutrality an equally large number of positive particles are required.

\begin{figure}
  \centerline{\psfig{file=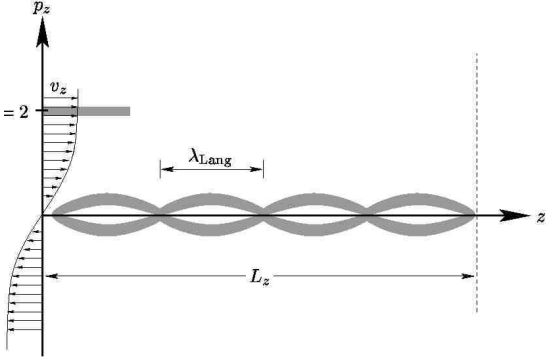,width=0.40\textwidth,clip=}}
  \caption{The initial configuration of electron phase space including
  the beam.  The initial disturbance of the background plasma is heavily
  exaggerated for reasons of better display.}
  \label{fig:Initial}
\end{figure}

  The initial condition is given by a homogeneous background plasma of
  density~$n_\ro{e}$, temperature~$\te$ and a longitudinal magnetic field
  $B_z = 0.1 \Unit{T}$.  This value is found to be sufficiently high to
  represent a much stronger field of 1000 \Unit{T}.  Initially there is no
  beam in the computational box.  The beam is injected later at the $z = 0$
  plane of the box with a density profile of $n \left( r \right) =
  n_\ro{b}/\left(1 + \exp \left[ \frac{ r - R }{\Delta R} \right]\right)$.
  Here $r$ denotes the distance from the $x$--$y$ center of the box and $R
  = 20 \Unit{m}$ and $\Delta R = 3 \Unit{m}$ give the radius and the
  sharpness of the electron beam.  The homogeneous background plasma is
  disturbed in phase space~(see Fig.~\ref{fig:Initial})
  in order to accelerate the excitation of Langmuir waves by giving the
  system a hint of the relevant length scales.  We used the wave length of
  the fastest growing Langmuir mode, which follows from an analytic
  calculation of the linear growth rate.  The disturbance is chosen not to
  induce any finite charge or current densities, thus leaving the
  background plasma initially homogeneous, neutral and current free.

  The boundary conditions during the simulation are quite simple.  In the
  transversal directions $x$ and $y$ we chose periodic boundaries and in
  the longitudinal direction~$z$ the box is open, which means that plasma
  particles and fields can leave the box unaffected.  In addition at the
  lower $z$ plane~($z = 0$) an electron component is continuously injected
  with the density profile mentioned above, a mean $\gamma\beta$ of 2 and
  the temperature~$\tb$.

  The simulation runs for 600 time steps with $\delta t = 2.41 \Unit{ns}$.
  The total time simulated was $1.44 \Unit{\mu s}$, after which the growing
  nonlinear Langmuir waves saturate and a quasi stationary situation had
  developed as it is shown in the next section.

\section{Results}
\begin{figure*}
  \centerline{\psfig{file=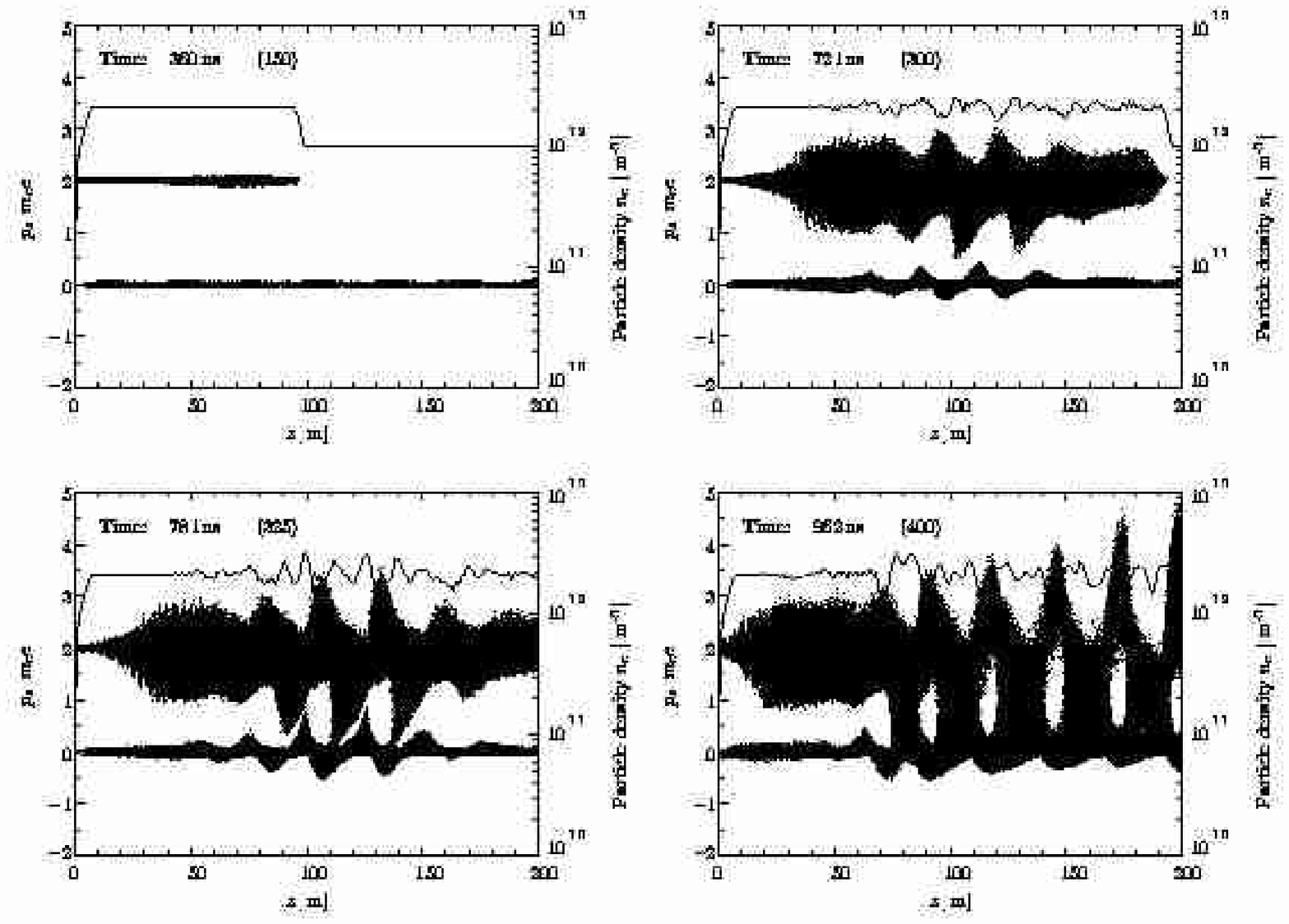,width=0.99\textwidth,clip=}}
  \caption{Electron phase space in $z$--direction after 150, 300, 325 and
  400 timesteps.  The quasi particles are plotted as dots according to
  their corresponding position in $z$~(ordinate) and their
  momentum~$p_z$~(left abscissa).  The solid curves present the particle
  density of the electrons~(right abscissa).}
  \label{fig:Phasespace}
  \centerline{\psfig{file=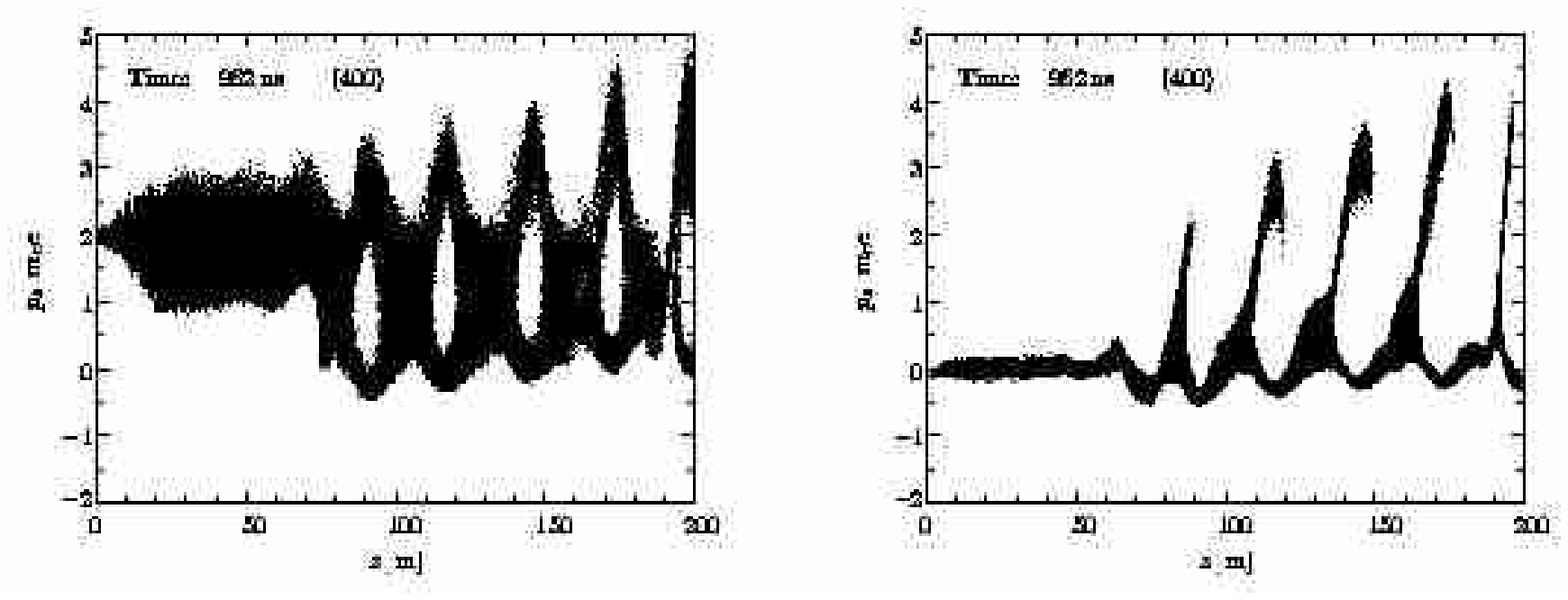,width=0.99\textwidth,clip=}}
  \caption{Electron phase space in $z$--direction after 400 timesteps
  separated for beam~(left) and background~(right) electrons.  }
  \label{fig:Phasespace_sep}
\end{figure*}

\begin{figure*}
  \centerline{\psfig{file=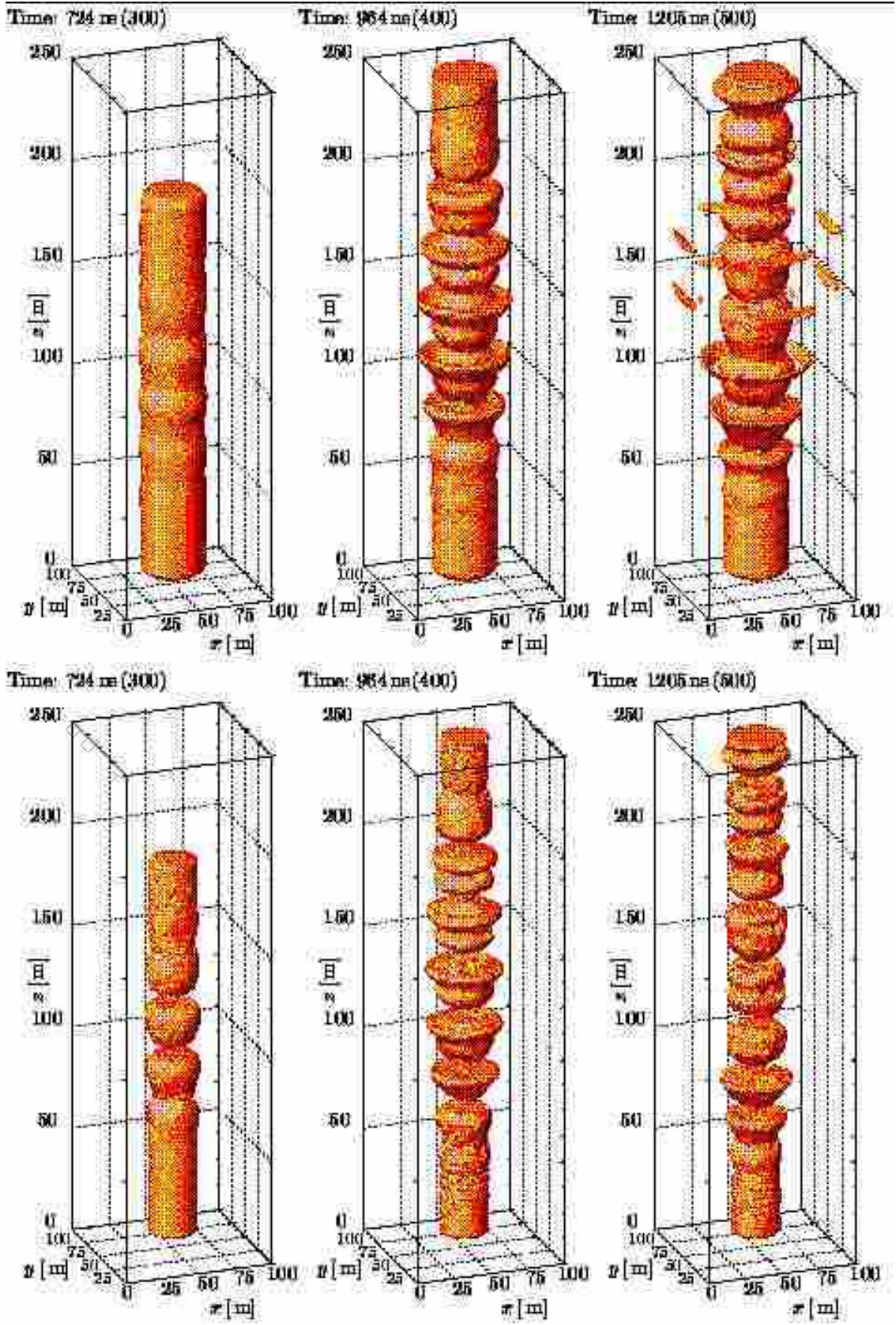,width=0.83\textwidth,clip=}}
  \caption{ Three dimensional contour plots of the electron particle
     density.  For the timesteps~300, 400 and 500~(left to right) the
     surfaces are shown, where the density is~$n = 1.5 \cdot 10^{12}
     \Unit{m}^{\sm 3}$~(top) and $n_\ro{e} = 1.9 \cdot 10^{12}
     \Unit{m}^{\sm 3}$~(bottom) respectively.  }
  \label{fig:Density_shd.eps}
\end{figure*}

\begin{figure*}
  \centerline{\psfig{file=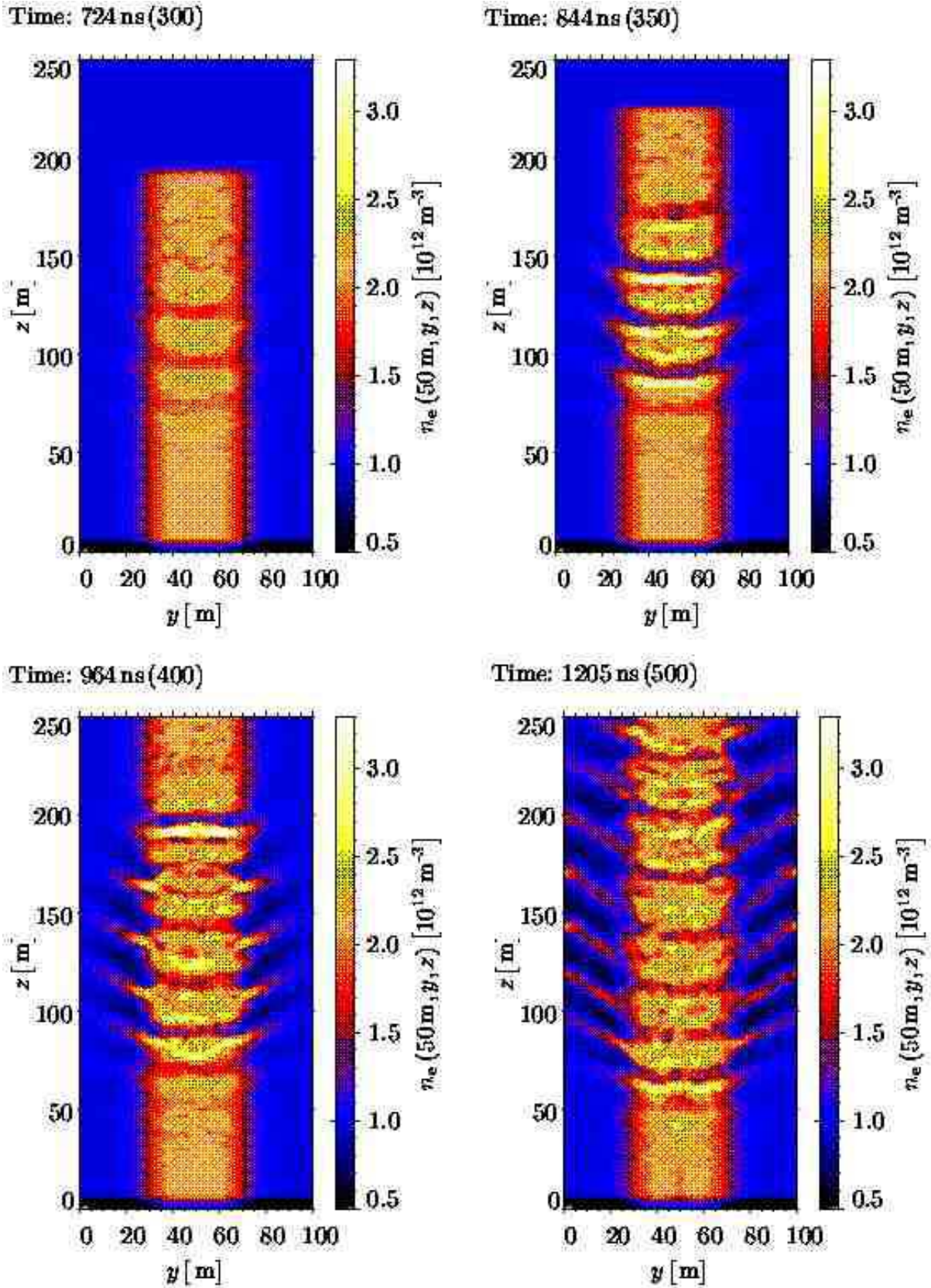,width=0.90\textwidth,clip=}}
  \caption{two dimensional contour plots of the electron particle density at
  timesteps 300,350,400 and 500.  Shown is a cut through the center of the
  penetrating electron beam at~$x = 50 \Unit{m}$.  }
  \label{fig:Density_con.eps}
\end{figure*}

  The beam particles entering the computational box interact with the faint
  traces of fluctuations we introduced in our initial setup and excite the
  appropriate modes to grow.  In Fig.~\ref{fig:Phasespace} on
  page~\pageref{fig:Phasespace} the temporal evolution of the electrons
  phase space in $z$--direction is presented by showing phase diagrams for
  the time steps 150, 300, 325 and 400.  In the first plot~(upper left) the
  penetrating electron beam can be seen together with the background
  plasma, both are still nearly undisturbed.  In the following plots the
  development of strong Langmuir waves can be observed.  After 781
  \Unit{ns}, the disturbance in the beam and the background has grown so
  strong, that both components start to mix, i.e. there is no distinction
  between beam and background electrons anymore.  Beam electrons are
  strongly decelerated even until $v\simeq 0$, whereas background electrons
  are accelerated up to a Lorenz factor of 5.  This can be seen best in
  Fig.~\ref{fig:Phasespace_sep} on page`\pageref{fig:Phasespace_sep}, where
  each of the both components are plotted in separate diagrams for the
  time step 400.  In Fig.~\ref{fig:Phasespace_sep}~left the phase space of
  the beam electrons is shown.  Clearly visible are the ring structures of
  fully developed Langmuir waves.  Some of beam electrons are accelerated
  up to a $\gamma\beta$ of 4--5 as it is the case for the background
  electrons, which is shown in Fig.~\ref{fig:Phasespace_sep}~right.  On the
  other hand the beam electrons are also decelerated to zero momentum
  during their interaction with the Langmuir waves.  The extremely strong
  accelerations are due to the strong electrostatic fields of the excited
  Langmuir waves.  Since the Langmuir turbulence structure in
  Fig.~\ref{fig:Phasespace} moves with a much lower speed~(which is even
  continuously decreasing throughout the simulation) than the yet unaffected
  beam electrons, it acts as the wiggler field that stimulates the
  \acro{CI CS}-emission.  The electrostatic fluctuations directly correspond
  to electron density fluctuations~(solid curve in
  Fig.~\ref{fig:Phasespace}), which reach a value of $\sim +100\%/-50\%$
  oriented at $n = n_\ro{e} + n_\ro{b} = 2 \cdot 10^{12} \Unit{m}^{\sm 3}$.
  The structure of the electron density is better seen in
  Fig.~\ref{fig:Density_shd.eps}~(page \pageref{fig:Density_shd.eps}) and
  Fig.~\ref{fig:Density_con.eps}~(page \pageref{fig:Density_con.eps}),
  where the temporal evolution of contour surfaces at $n = 1.5 \cdot
  10^{12} \Unit{m}^{\sm 3}$ and $n_\ro{e} = 1.9 \cdot 10^{12} \Unit{m}^{\sm
  3}$ and of a cut through the electron beam is shown.  The electron
  density varies between $4$~and~$1$ times $10^{12} \Unit{m}^{\sm 3}$,
  which implies the strong electrostatic fields.  It can be clearly seen,
  that the beam decays in a series of "pancakes".  Such pancakes are
  absolutely necessary to explain the coherent nature of the pulsed radio
  emission of neutron stars.  Why? Since the emission from relativistic
  particles is confined to a forward cone with half angle $\sim 1/\gamma$
  and the emission is nearly along the magnetic field.  Thus, in Fourier
  space bunching emission corresponds to the component $k_\perp$ to the
  fields being smaller than the component $k_\parallel$ along the field
  line by a factor $1/\gamma$.  This corresponds to flat pancake shaped
  bunch with the normal within an angle $1/\gamma$ of~$B$
  \cite{Melrose2000}.  This is exactly what we observe in our simulations!
  We briefly note that our simulations are the first ones which show this
  kind of pancake bunches!  The number of particles in such a pancake
  corresponds to the the maximum coherence achievable for \acro{CICS} and
  it should be noted, that in our simulation the thickness of such a
  pancake is significantly larger~($\gtrsim 1 \Unit{m}$) than the Debye
  length which is usually used in analytical models.  A typical volume of
  one bunch is about $10 m^3$ giving a total number of particles per bunch
  of about $10^{13}$.  Our simulations prove, that a much higher coherence
  is achievable than expected from analytical models.

\begin{figure}
  \centerline{\psfig{file=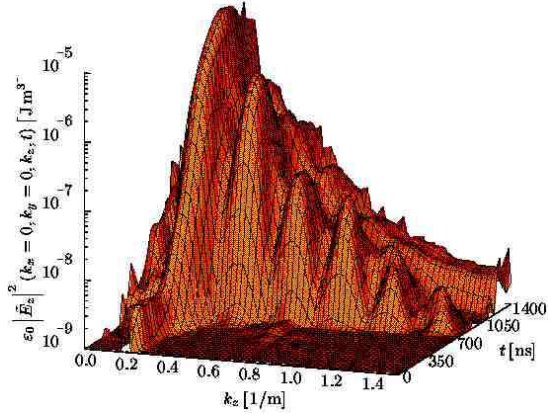,width=0.40\textwidth,clip=}}
  \caption{The electrostatic mode energy density is plotted against
  wavenumber~$k$ and time~$t$.  }
  \label{fig:Growth}
\end{figure}

  The growth of Langmuir waves can be deduced by the Fourier transform of
  the electrostatic field~$\tilde{E}_z$.  In Fig.~\ref{fig:Growth}
  the energy density of the electrostatic modes is plotted versus the
  corresponding wavenumber~$k$ and time~$t$.  The $k \simeq 0.25
  \Unit{m}^{\sm 1}$ mode is growing by four orders of magnitudes within
  some ten nanoseconds.  It should be noted however, that the fastest
  growing mode is slightly off compared to the initially given wave number
  can be seen in Fig.~\ref{fig:Growth} as the small peak at early times.
  This means, that for our simulation the analytical value for the fastest
  growing mode is not precisely correct and it proves, that although the
  initial disturbance has given a hint to the system it has not forced it
  towards an unphysical solution.  We note that the wave energy density is
  distributed to higher $k$ when it exceeds a value of about $10^{\sm 7}
  \Unit{J}\Unit{m}^3$, driven by nonlinear wave--wave interactions.

\begin{figure}
  \centerline{\psfig{file=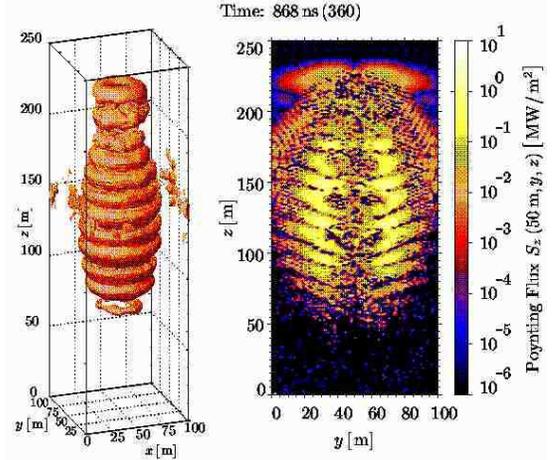,width=0.4\textwidth,clip=}}
  \caption{Structure of the emitted pointing flux as 3D contour surface of
  the value~$2.5 \cdot 10^4 \Unit{W}/\Unit{m}^2$~(left) and as a cut
  through the center of the beam at $x = 50 \Unit{m}$~(right). }
  \label{fig:Poyntingstrucure}
\end{figure}

\begin{figure}
  \centerline{\psfig{file=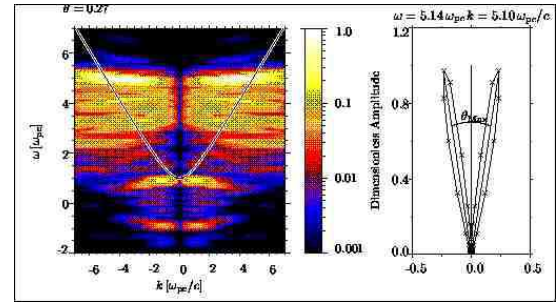,width=0.4\textwidth,clip=}}
  \caption{Dispersion relation~(left) and emission characteristic~(right)
  of the emitted radiation.  For the left plot the angle $\theta_\ro{M}$,
  the angle of maximum intensity, has been used.  The curve in the left
  plot gives the dispersion branch of electromagnetic radiation in a
  plasma.  The right diagram presents the intensity of the radiation for a
  given $\theta$ in a polar plot.  $\theta$ is the angle between beam
  direction and direction of emission.  The values in both diagrams are
  normalized to a value of one.}
  \label{fig:Dispersion}
\end{figure}

\begin{figure}
  \centerline{\psfig{file=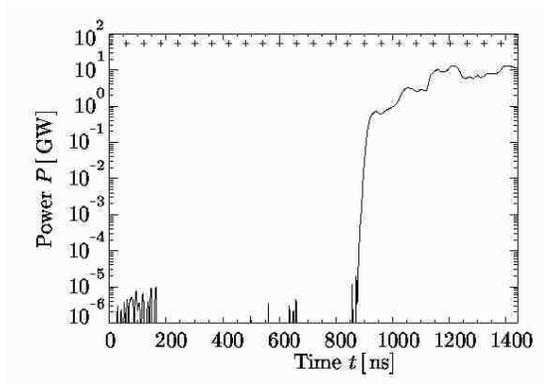,width=0.40\textwidth,clip=}}
  \caption{Total integrated poynting flux leaving the computational box at
  the top~($z = 250 \Unit{m}$) against time.  The $+$~symbols show the
  total power injected as kinetic energy of the beam electrons.  }
  \label{fig:Poyningtotal}
\end{figure}

  It is important to recognize that the generated electrostatic
  fluctuations react on the beam electrons similar to the wiggler fields in
  the laboratory \acro{FEL}, with one major difference.  In our case the
  wave wiggling self-consistently excited is along the propagation
  direction of the beam, whereas in the \acro{FEL}-case the external
  magnetic wiggling is perpendicular to the beam propagation.  The strongest
  emission originates in regions with the largest density gradients
  Fig.~\ref{fig:Poyntingstrucure}.
  Every de-- and acceleration region forces the beam electrons to emit
  Hertz' dipole radiation, which is beamed in forward direction, as can be
  seen in Fig.~\ref{fig:Dispersion}~(right).
  $\theta_\ro{M}$ represents the angle of maximum intensity, expected from
  theoretical consideration~($\tan \theta_\ro{M} = 1/2\gamma\beta$) with a
  $\gamma\beta = 2$.  The striking resemblance with a relativistically
  beamed Hertz' dipole is obvious, but the emission characteristic in our
  simulation also gives precisely the expected opening angle of the
  emission cone, which is proposed for a \acro{CICS}-process.  In
  Fig.~\ref{fig:Dispersion}~(left) the "spectrum" of the emitted radiation
  is shown, which has a strong peak at $\omega = 5 \wpe$ and a couple of
  minor peaks at $4 \wpe$ and $3 \wpe$ due to nonlinear wave coupling.
  Again we find rather precisely the theoretically expected value of $w =
  \gamma^2 \wpe$.  The total power emitted by the simulated computational
  box is shown in Fig.~\ref{fig:Poyningtotal}.
  It is the escaping poynting flux at the top of the computational box~($z
  = 250 \Unit{m}$) integrated over the top surface.  After about $850
  \Unit{ns}$ the emitted power rises within some ten nanoseconds by more
  than $6$ orders of magnitude to a value of $1 \Unit{GW}$.  Afterwards the
  total output power increases more slowly by one more order of magnitude
  and reaches a maximum of $10 \Unit{GW}$ continuous radiative power at the
  end of the simulation.  This is a significant fraction of the power
  injected by kinetic energy of the beam particles, shown in
  Fig.~\ref{fig:Poyningtotal} as $+$~symbols at a constant value of $50
  \Unit{GW}$.  The rest of the energy leaves the box as kinetic energy of
  beam electrons, which would probably be also emitted if the computational
  would have a greater size in z-direction.

\section{Discussion}
  We have shown that in a strongly magnetized plasma a relativistic
  electron beam can be forced to emit highly coherent radio emission by
  self-induced nonlinear density fluctuations.  Such slowly moving nonlinear
  structures oscillate with the local plasma frequency at which the
  relativistic electrons are scattered.  Beam electrons dissipate a
  significant amount of their kinetic energy by inverse Compton radiation
  at a frequency of about $\gamma^2\omega_{\rm pe}$.  Since the beam is
  sliced into pancake structures which experience the same electric field
  the inverse Compton scattering is coherent.  Such a process is a very
  promising candidate for the coherent radio emission of pulsars.

\section*{Acknowledgments}
\begin{acknowledgements}
  The authors would like to thank the John von Neumann Institute for
  Computing~(\acro{NIC}) and the Leibniz Rechenzentrum M\"unchen for the
  granted computational time and the Deutsche Forschungsgemeinschaft which
  supported our work through the grant LE~1039/3-1.
\end{acknowledgements}

%


\clearpage

\end{document}